\documentclass{cimento}
    \title{Study of Systematic Uncertainties of Single Top Production at ATLAS}
    \author{ G.~Khoriauli\from{ins:bonn}}
    \instlist{\inst{ins:bonn}Physikalisches Institut, Universit\"at Bonn, Germany}
    \PACSes{\PACSit{14.65.Ha}{Top quarks}}

\usepackage{graphicx}
\usepackage{mathptmx}

\begin{document}

\maketitle

\begin{abstract}
Sytematic uncertainties to the single top production cross section measurement at the ATLAS experiment has been studied. 
Different sources of systematic uncertainties such as detector luminosity, jet energy calibration, SM background 
normalization, PDF parameterization and others have been considered. Large scale Monte-Carlo events simulation has been 
performed to estimate the contribution of each source in the overall uncertainty. The study was done for cut based 
analysis as well as for multivariate analysis of the single top measurements in ATLAS.  The total systematic 
uncertainties of the single top cross-section measurements in its three production channels have been estimated 
at 1 $\mathrm{fb}^{-1}$ and 10 $\mathrm{fb}^{-1}$ integrated luminosity of the LHC.
\end{abstract}

\section{Introduction}
The single top production at the LHC in the three different channels ($t$, $s$ and $Wt$) is expected to have a sizable
production rate. 
This should allow to accurately measure these weak processes and determine the cross section in the different channels~\cite{markus}. 
As the detailed cut based and multivariate analysis show, the single top production processes in ATLAS could be discovered at the level 
of few hundreds of inverse pb (depending on the channel) of the integrated luminosity of LHC~\cite{markus}. 

However, several anticipated sources of systematic uncertainties will significantly affect the precision 
of this measurement. Detailed study of the systematic uncertainty sources has been performed within the ATLAS top physics working group. 
Influence of the uncertainties of the LHC luminosity, trigger efficiencies, jet energy calibration, b-jet tagging efficiencies, 
lepton identification, background normalization, initial and final state radiation (ISR/FSR), parton distribution functions (PDF) and
b-fragmentation parameterizations on the single top cross section measurements
were estimated using detailed Monte-Carlo studies of the processes in the ATLAS detector. Systematics were determined for cut based and 
multivariate analyses at different reference integrated luminosities. Estimation of the systematics from some
uncertainty sources would have required very large statistics Monte-Carlo simulation. In those cases a re-weighting method was implemented, 
validated and used for estimation of the uncertainties. The re-weighting approach was necessary for PDF and b-fragmentation parametrization
uncertainty studies.

\section{Parton distribution functions}
The PDF parametrization is based on a fit to a large number of experimental data points and inherits uncertainties from the data used in the 
fit. Then the PDF parametrization uncertainty can propagate to the uncertainty of the measured physical observable, for example, of the cross section. 
In order to estimate the resulting PDF uncertainties on the single top cross section measurements, first the PDF uncertainty on 
the selection efficiencies of the signal and main background ($t\bar{t}$) has been estimated. Then, the selection efficiencies were translated to
uncertainties in the different single top channel cross section measurements. PDF uncertainties were estimated at the NLO level. For this, the 
MC@NLO Monte-Carlo generator together with the CTEQ6.1 NLO PDF error set has been used for the event generation 
(for the Wt-channel AcerMC generator has been used instead).
Events were generated at the so called 
central value PDF (corresponding to the PDF parameterization fit parameters at the global minimum of the fit) of the CTEQ6.1 ~\cite{CTEQ61} set. 
Each event
was assigned a set of weights determined as follows, 

\begin{eqnarray*}
w^{\pm}_{i} = \frac{f_{1}(x_{1},Q;S^{\pm}_{i}) \cdot f_{2}(x_{2},Q;S^{\pm}_{i})}{f_{1}(x_{1},Q;S_{0}) \cdot f_{2}(x_{2},Q;S_{0})}
\end{eqnarray*}

Here, $f_{1}$ and $f_{2}$ are PDF values for a given hard scattering process (characterized by flavors $f$ and momentum fractions $x$ of 
the initial partons, and by $Q$, the event scale) evaluated for the $i^{th}$ error PDF pair $S^{\pm}_{i}$. The terms in the denominator 
are obtained with the central value PDF. For the uncertainty calculation, events were taken into account together with the corresponding 
PDF weights. This means that weighted and normalized sums of selected events were determined as given by the formula: 

\begin{eqnarray*}
W^{\pm}_{i} = \frac{\sum_{j=1}^{m}{w^{\pm}_{ij}}}{\sum_{k=1}^{M}{w^{\pm}_{ik}}}
\end{eqnarray*}

where, $M$ and $m$ are the total and selected numbers of single top events respectively. The $W^{\pm}_{i}$'s form a set of normalized $2N$ 
weight sums, which replace selection efficiencies $\epsilon^{max}_{i} \rightarrow \textrm{max}(W^{+}_{i}, W^{-}_{i})$ and 
$\epsilon^{min}_{i} \rightarrow \textrm{min}(W^{+}_{i}, W^{-}_{i})$ in the master equation for the up/down systematic uncertainties due to PDF,  

\begin{eqnarray*}
\Delta^{+} \epsilon = \sqrt{ \sum_{i=1}^{N}{ \textrm{max}(\epsilon^{max}_{i} - \epsilon^{0}, 0)^{2} }} \qquad\mbox{and}\qquad
\Delta^{-} \epsilon = \sqrt{ \sum_{i=1}^{N}{ \textrm{max}(\epsilon^{0} - \epsilon^{min}_{i}, 0)^{2} }}. 
\end{eqnarray*}

In these equations, $N$ is the number of global fit parameters and $\epsilon^{0}$ is evaluated 
at the central value PDF. Figure 1 shows the central value corresponding selection efficiency for the t-channel and for $t\bar{t}$
(applying the t-channel selection) in the $e^{\pm}$ - final states and the variation of the re-weighted selection efficiencies.

\begin{figure}[hbt]
\begin{center}
	\includegraphics[height=4.0cm]{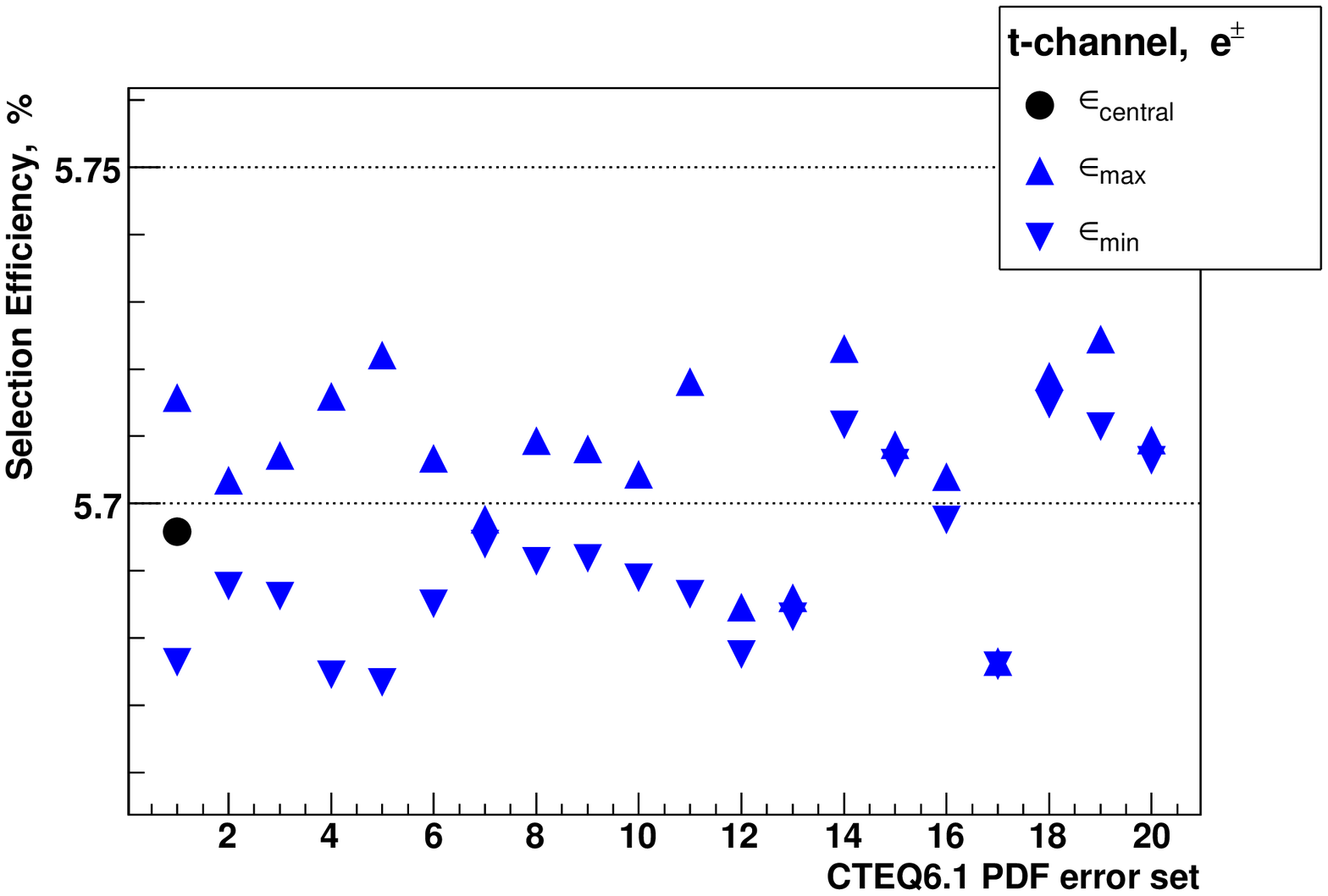}
	\includegraphics[height=4.0cm]{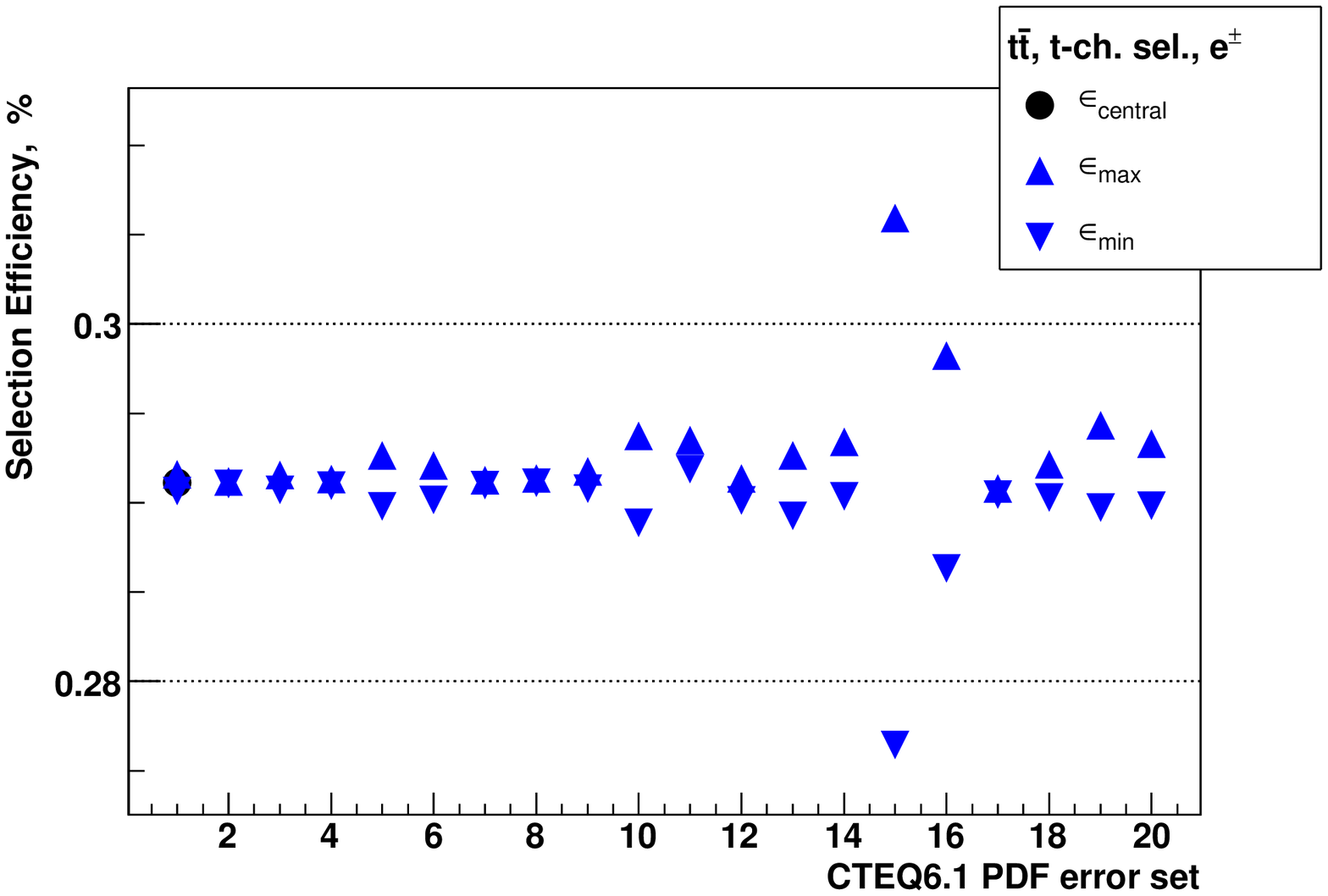}
	\caption{\small{\it Variation of PDF re-weighted selection efficiencies. Left: t-channel into $e^{\pm}$ - final state; 
                            right: $t\bar{t}$ (with t-channel selection) into $e^{\pm}$ - final state}}
\end{center}
\end{figure}

\section{b-quark fragmentation}
The b-quark fragmentation determines the spectrum of the b-flavored hadron distribution as a parameterized function of a 
Lorenz-invariant energy-momentum variable - $z$. 
The b-fragmentation parametrization uncertainty affects the b-tagging efficiency and the b-hadron energy spectrum in Monte-Carlo simulation 
and therefore, it is an additional source of systematic uncertainty in the single top selection efficiency. The uncertainty was 
determined as a difference between the single top selection efficiencies due to the choice of the different b-fragmentation models,
the Peterson~\cite{Peterson} 
compared to the Lund-Bowler~\cite{Lund, Bowler} model. 
The latter leads to an increase of the expected b-tagging efficiency (0.6) by 0.011.
This result was used to estimate the corresponding selection efficiency uncertainty using an event weighting method. The normalized 
weight sum for selected events was determined considering the specific selection cut requirements on b-jets in the different single top channels:
 
\begin{eqnarray*}
w^{\pm}(\mathrm{s-channel}) = \frac{1}{N} * \sum_{i=1}^{n} { \frac{P_{i}^{0.6 \pm 0.011}(\textrm{exactly 2 truth b-jets out of M with } p_{T}>30 \textrm{GeV are tagged})} {P_{i}^{0.6}(\textrm{exactly 2 truth b-jets out of M with } p_{T}>30 \textrm{GeV are tagged})} }
\end{eqnarray*}
\begin{eqnarray*}
w^{\pm}(\mathrm{t-channel}) = \frac{1}{N} * \sum_{i=1}^{n}{ \frac{P_{i}^{0.6 \pm 0.011}(\textrm{at least 1 truth b-jet out of M with } p_{T}>50 \textrm{GeV is tagged})} {P_{i}^{0.6}(\textrm{at least 1 truth b-jet out of M with } p_{T}>50 \textrm{GeV is tagged})} }
\end{eqnarray*}

where, $P_{i}^{0.6}$ and $P_{i}^{0.6 \pm 0.011}$ are the binomial probabilities for the $i$-th event calculated at the b-tagging efficiency values: 
0.6, 0.611 and 0.589. $M$ is the number of true b-jets in the $i$-th event satisfying a $p_{T}$ requirement. $N$ and $n$ are the number of total
and selected events respectively. The accumulated normalized weight sums were used to calculate the corresponding up and down uncertainties on 
the selection efficiency. This is done using a similar procedure to the PDF uncertainty calculations. 

\section{Estimation of the total uncertainty}
The results of the PDF, b-fragmentation and all other systematic uncertainties were used to determine an overall uncertainty 
of the cross section measurement using the master formula,
$\sigma = { N_{tot} - B \over a \times {\cal{L}} }$;
Where, $N_{tot}$ is the sum of signal and background events provided by the Monte Carlo simulation, $B$ is the sum of all background contributions, 
$\epsilon$ is the signal selection efficiency and ${\cal{L}}$ is the luminosity. 
The effect of each source of uncertainty on these variables were combined and propagated to the measured cross section using a toy Monte Carlo method, 
which randomly generates $N_{tot}$ according to a Poisson distribution, and randomly varies $B$ and $\epsilon$ for every systematic source by 
an amount chosen around its central value, according to a Gaussian distribution. This procedure is performed a few thousand times and the RMS of 
the resulting distribution is interpreted as the total uncertainty.
The total uncertainty is dominated by systematic effects due the background normalization. 
Reducing the background contamination to increase $S/B$ would help to reduce the systematic uncertainty. 
Multivariate analyses were used to further optimize the selection beyond what can be achieved with cuts on the existing variables~\cite{markus}. 
Figure 2 shows the cross-section systematic uncertainties for the different channels at the 1 $\mathrm{fb}^{-1}$
and 10 $\mathrm{fb}^{-1}$ 
integrated luminosities of the LHC calculated for the multivariate analyses. Table 1 provides the statistical and systematic uncertainties 
for these channels for the above integrated luminosities.

\begin{figure}[t]
\begin{center}
	\includegraphics[height=4cm]{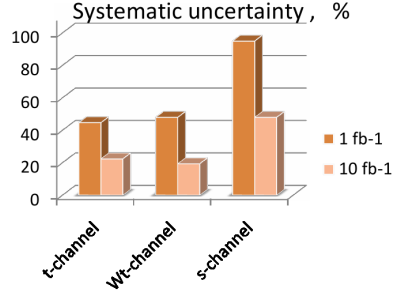}
	\caption{\small{\it Single top cross-section measurement systematic relative uncertainties}}
\end{center}
\end{figure}

\begin{table}[htb] 
 \begin{center} 
   \begin{tabular*}{0.66\textwidth}{l|cc||cc} 
   \hline 
	channel & \multicolumn{2}{c| |}{Analysis at 1 $\mathrm{fb}^{-1}$}   & \multicolumn{2}{c}{Analysis at 10
	$\mathrm{fb}^{-1}$}  \\[1mm]
	 	& $\pm$ stat. \% 	& $\pm$ syst. \%  	& $\pm$ stat. \% 	&  $\pm$ syst. \%  \\[1mm]
   \hline
   \hline
	t		&  5.7 &  21.7 &  1.8 &  9.8	\\[1mm]
   \hline
	s		&  64.0 &  95.0 &  20.0 &  48.0	\\[1mm]
   \hline
	Wt		&  20.6 &  48.0 &  6.6 &  19.4	\\[1mm]
   \hline
   \end{tabular*}
   \caption[]{\small{\it Cross section measurement relative uncertainties for the different channels of the single top production at ATLAS. 
		  Multivariate analysis results}}
 \end{center} 
\end{table}

\end{document}